\documentclass[12pt]{article}

\author{{\small Colin Goldblatt} \\ {\small School of Earth and Ocean Sciences, University of Victoria, BC, Canada.}\\{\small\url{czg@uvic.ca}}}
\title{Atmospheric Evolution}
\date{}

\usepackage{natbib}
\usepackage[version=3]{mhchem}
\usepackage{amssymb}
\usepackage{graphicx}
\usepackage{url}

\newcommand{\dc}{\ensuremath{^{\circ}}C}
\newcommand{\wmm}{W\,m$^{-2}$}

\begin{document}

\maketitle






\textsc{Definition}---Earth's atmosphere has evolved as volatile species cycle between the atmosphere, ocean, biomass and the solid Earth. The geochemical, biological and astrophysical processes that control atmospheric evolution are reviewed from an ``Earth Systems'' perspective, with a view not only to understanding the history of Earth, but also to generalizing to other solar system planets and exoplanets.


\section{Introduction}

The evolution of Earth's atmosphere has been a function of the cycling of volatile species between the atmosphere, ocean, biomass and the solid Earth. The composition of the atmosphere largely determines the climate, thus whether life can thrive, and the rate of many geochemical processes. Atmospheric evolution is an inherently interdisciplinary problem, requiring an ``Earth system'' approach; it is replete with feedback processes and neither a single branch of science nor study of any single part of the Earth is sufficient to make progress. 

In the preface to \emph{The Biosphere}, a visionary work of what would now be Earth system science, \citet{vernadsky-26} wrote\newline
``Historically, geology has been viewed as a collection of events derived from insignificant causes, a string of accidents. This of course ignores the scientific idea that geological events are planetary phenomena, and that the laws governing these events are not peculiar to Earth alone.''\newline
This review proceeds from that paradigm, with a focus on process and mechanism applicable to the evolution of Earth's atmosphere that are generalisable to other atmospheres, either comparison planets in our solar system or other inhabited planets. 

Indeed, as we enter the age of exoplanets, understanding the evolution of Earth's atmosphere gains greater meaning. The practical method of remotely determining the geochemistry of a planet, and potentially detecting life, is spectroscopic analysis of the planetary atmosphere \citep{lov-65,hl-67}. Interpreting whether a remotely detected atmospheric composition implies life requires an understanding the range of compositions that an inhabited planet atmosphere could have, and Earth history is the biggest sampling of this parameter space available. 

\begin{figure}
\begin{center}
\includegraphics[width=0.9\textwidth]{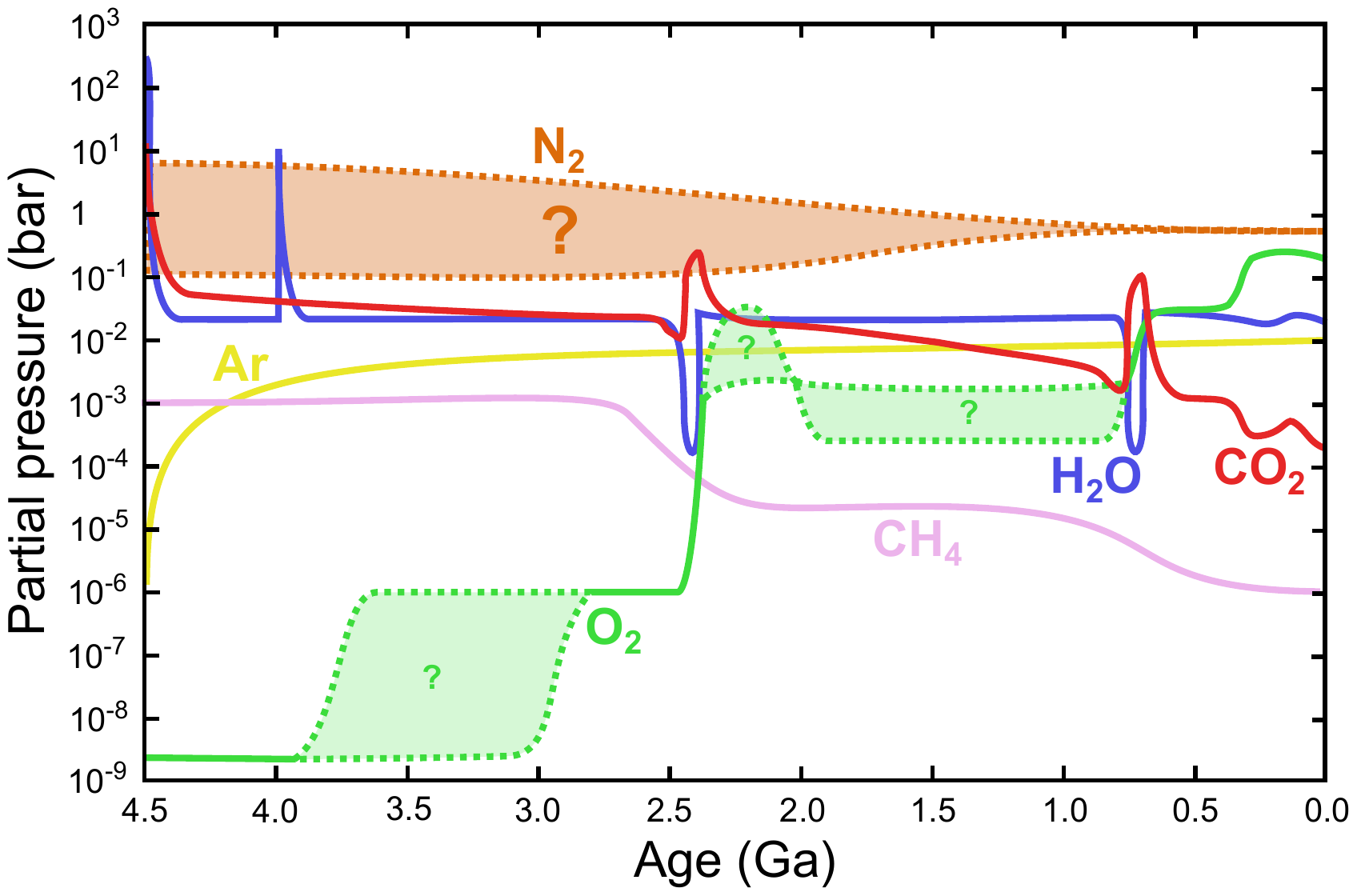}
\end{center}
\caption{The evolution of Earth's atmosphere (an artist's conception). Uncertainties of a factor of a few are implied; smooth lines mostly correspond to  lack of theory and constraints. Snowball Earth periods likely involve two or three glacials and interglacials, but are shown as one for simplicity. } \label{f-history}
\end{figure}

The structure of this review is to first introduce the role of the atmosphere in maintaining climate, then treat atmospheric gasses beginning with the most abundant: nitrogen, oxygen, carbon dioxide, then water. Trace species are treated with the more abundant gas that they are most closely linked to. Figure \ref{f-history} gives a schematic summary of atmospheric evolution. 

\section{The atmosphere and climate}
\subsection{ A climate primer}

On timescales relevant to atmospheric evolution, Earth's atmosphere is in energy balance: energy in equals energy out. Solar energy incident on Earth ($F_\mathrm{solar}$) dominates over the transfer of accretional and radiogenic heat to the surface by four orders of magnitude (342\,Wm$^{-2}$ versus 82\,mWm$^{-2}$), so the Sun is the driver of climate. A fraction of the incident sunlight is reflected back to space by clouds, molecules or the surface, referred to as the plantary albedo ($\alpha_p$), around 0.3 for Earth today. Some sunlight is absorbed by water vapour in the lower atmosphere, but the bulk is absorbed at the surface. The result is that the atmosphere is heated from the base. Convection therefore dominates the lowest layer of the atmosphere, the troposphere, with temperature decreasing with height as rising air expands and cools. Above the top of this layer (the tropopause), vertical motions are minimal and the temperature structure is set by balance of radiative absorption and emission. This layer is the stratosphere. Today, temperature increases in the stratosphere (a ``temperature inversion'') due to strong absorption of UV light by ozone.

Were Earth air-less, radiative emission to space would equal surface emission and planetary temperature would be trivial to calculate. Surface emission approximates a black-body and would balance absorbed solar radiation averaged over the surface:  $(1-\alpha_p)F_\mathrm{solar} = \sigma T_\mathrm{surf}^4$, where $\sigma$ is the Stefan-Boltzmann constant and $T_\mathrm{surf}$ is the surface temperature, giving $T_\mathrm{surf} = $\,K.

Earth, of course, has an atmosphere which imparts a greenhouse effect, meaning that thermal emission to space for a given surface temperature is less than the black-body flux. This greenhouse effect occurs as infra-red radiation is absorbed in the atmosphere by gases (``greenhouse gases'') or particles (e.g. clouds), which then emit radiation both upward and downward. Given that the atmosphere is colder than the surface, the total thermal radiation to space is less in the presence of greenhouse absorption than it would be without. 

\begin{figure}
\begin{center}
\includegraphics[width=0.7\textwidth]{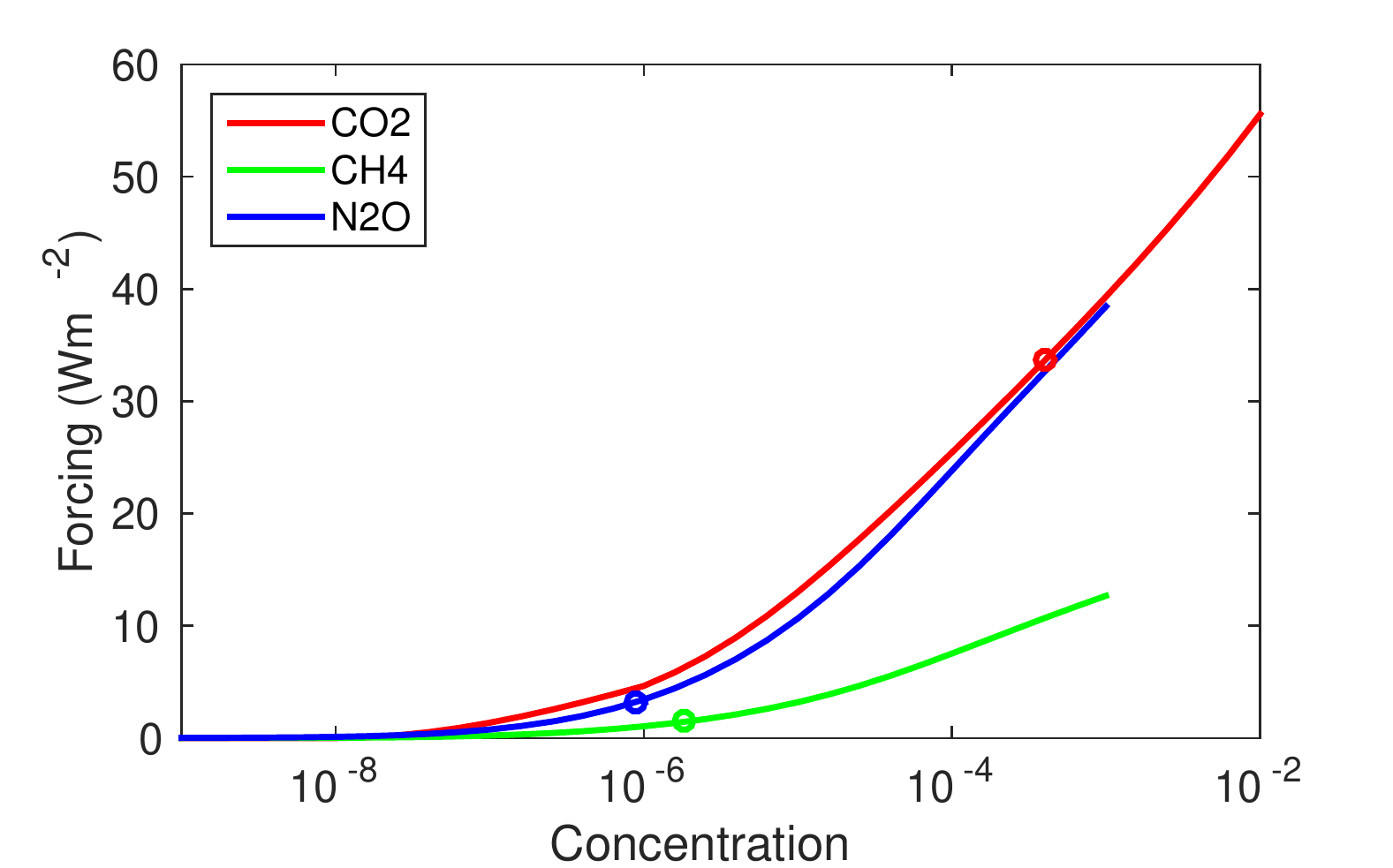}
\end{center}
\caption{Greenhouse gas forcing (change in net flux at the tropopause) relative to zero concentration of that gas \citep{Byrne2014grl}. }\label{fig:ghforcing}
\end{figure}
For greenhouse gases at climatically relevant levels, it is fundamental to note that the strength of the greenhouse effect is proportional to the logarithm of gas abundance (Figure \ref{fig:ghforcing}). This relates to the absorption spectra of real gasses; see an excellent review of the greenhouse effect by \citep{Pierrehumbert2011} for an explanation. Arising from this is the common misunderstanding that methane is an inherently strong greenhouse gas. It is actually a rather weak infrared absorber, however its low abundance \emph{in the present day atmosphere}, 300-times lower than carbon dioxide, means that adding a fixed number of moles of carbon as \ce{CH4} rather than \ce{CO2} will give a larger forcing. 

\subsection{The Faint Young Sun Paradox}

The amount of solar energy incident on Earth has increased by 30\% through Earth history due to stellar evolution, yet it does not appear that Earth was substantially colder in the past \citep{sm-72}. This is the so-called Faint Young Sun Paradox (FYSP), which substantially motivates the study of atmospheric evolution from a palaeoclimate perspective: the past atmospheric composition must have been very different in order to keep Earth warm. 

Primary evidence for temperate early Earth climate comes from the ubiquity of fluvial successions and sub-aqueously deposited strata in the Archean, compared to a dearth of glacial features \citep{nisbet-87}. Whereas regional glaciation recurs every $\sim\!100$\,Myr during the Phanerozoic, Precambrian glaciation is rare: low-latitude ``Snowball Earth'' glaciation in the Cryogenian (Neoproterozoic) and Siderian (Palaeoproterozoic) and regional glaciations in the Ediacaran (Neoproterozoic) and Mesoarchean \citep[][Paul Hoffman, private communication]{evans-03}.

Given lower insolation, temperate climate requires either a much stronger greenhouse effect or a lower planetary albedo, or both. Most attention has focussed on greenhouse gases, with likely contributions from carbon dioxide \citep{ocr-79} and methane \citep{hdkk-08} and many other possible contributors \citep{Byrne2014}.  Clouds could contribute to a stronger greenhouse effect if there were more high clouds, or reduce the  albedo if there were less low clouds \citep{gz-11}. 

See \citet{Feulner2012} for a recent review of the FYSP.

\section{Nitrogen}

Nitrogen is present in the atmosphere as di-nitrogen (\ce{N2}), the dominant background gas, and as more reactive and radiatively active species, primarily nitrous oxide (\ce{N2O}) and ammonia (\ce{NH3}). Nitrogen is a major element in life, and thus has a fast biological cycle. There is also a slow geological cycle which determines atmospheric \ce{N2} levels, which has only recently begun to be elucidated.

An understanding of the biological N cycle [CHAPTER-REF Nitrogen Cycle] underpins all theory on N evolution. In summary: for nitrogen to be bio-available, it must be chemically reduced (fixed) from \ce{N2} to ammonium (\ce{NH4^+}). Fixation is dominated by organisms; there is metaboloic cost to N-fixation, but this is selected for when N is a limiting nutrient. In oxic conditions (in confined oxygen oases in the Neoarchean, then widespread from the Palaoproterozoic onwards) \ce{NH4^+} may be returned to \ce{N2} via nitrification and denitrification. If denitrification is incomplete, nitrous oxide (\ce{N2O}) is produced, then returned to \ce{N2} by atmospheric photolysis. Under widespread anoxia (Meoarchean and earlier, and often in the Neoarchean), the major destruction pathway for (\ce{NH4^+}) would be outgassing of ammonia  from the ocean followed by atmospheric photolysis [CHAPTER-REF Atmospheric Chemistry]. Evolution of the N cycle is reviewed by \citet{Stueken2016}.

\subsubsection{Di-nitrogen}
\begin{table}
\begin{center}
\begin{tabular}{ll}
\hline
\textbf{Reservoir}  & \textbf{Size ($10^{18}$\,kg\,N)} \\
\hline
\textbf{Biomass} (live and dead)   & $\mathbf{9.3 \times 10^{-4}}$ \\
   \quad Land & $1.3 \times 10^{-4}$ \\
   \quad Ocean & $8.0 \times 10^{-4}$  \\
\textbf{Ocean} &  $\mathbf{2.4 \times 10^{-2}}$ \\  
\textbf{Atmosphere} &  $\mathbf{4.0}$ \\   
\textbf{Ocean crust} &  $0.47\pm$ \\ 
   \quad Sedimentary  & $0.41\pm 0.20$ \\
   \quad Igneous (altered) & $0.061\pm 0.007$ \\
\textbf{Continental Crust} & $\mathbf{2.0 \pm 0.16}$\\
   \quad Upper Crust & \\
     \qquad Sedimentary & $0.76 \pm 0.06$\\
     \qquad Metamorphic & $0.72 \pm 0.13$\\
     \qquad Volcanic & $0.017 \pm 0.020$\\
     \qquad Felsic intrusive & $0.31 \pm 0.04$\\
     \qquad Mafic intrusive & $0.0036 \pm 0.0015$\\
   \quad Lower crust & $0.18 \pm 0.06$\\
\textbf{Mantle} & $\mathbf{24.2\pm 16}$\\
   \quad MORB Source & $7.2\pm 5.9$\\
   \quad High N & $17\pm15$\\
\textbf{TOTAL} & $\mathbf{30.2\pm 16}$\\   
  \hline 
\end{tabular}
\end{center}
\caption{Earth system nitrogen reservoirs, after \citet{johnson2015}. The ``high N'' mantle reservoir is quite uncertain. The core is likely a large N reservoir, but is excluded as it is not thought to exchange with the atmosphere after core-mantle segregation.}\label{t-nres}
\end{table}

Di-nitrogen dominates Earth's atmosphere ($p\ce{N2} = 0.78$\,bar at present) and is subject to slow geologic cycling, making variations in the \ce{N2} inventory likely. Until recently, atmospheric \ce{N2} had generally been thought of as constant, with the implicit assumption that the biological cycle was entirely closed. However, small leakage of fixed N from the ocean to the geosphere drives a whole Earth N cycle, just as organic carbon burial drives a large part of the coupled carbon-oxygen cycle.  Recent whole-Earth nitrogen budgets have shown that there is more nitrogen in the silicate-Earth than in the atmosphere, and that this nitrogen is necessarily of atmospheric origin and fixed biologically. This calls for a cycle of nitrogen between the atmosphere and the crust and mantle, operating over billions of years. Variations in the size of the atmospheric \ce{N2} inventory by a factor of two or three are expected, but are very poorly constrained at present \citep{Goldblatt2009n2,johnson2015}.

Table \ref{t-nres} shows Earth's nitrogen reservoirs. The total nitrogen inventory is consistent with a chondritic source with $\sim 10$\% retention. Nitrogen in the core is assumed to be inaccessible to the rest of the Earth system, leaving dominant exchangeable reservoirs to be the mantle (a few times the present atmospheric inventory), continental crust (around half the present atmospheric inventory) and the atmosphere itself. Understanding of the evolution of the atmospheric reservoir must, therefore, emerge from understanding the history of exchange with the crust and mantle. 

The most important way that nitrogen enters rocks is as \ce{NH4^+}, which substitutes well for potassium ions (\ce{K^+}) in silicate rocks, as they have the same change and a similar ionic radii. Organic nitrogen in sedimentary rock is an additional source, and this may be transferred to silicate-bound \ce{NH4^+} during hydrothermal alteration, diagenisis or metamorphosis. The geological N cycle and budget were recently reviewed by \citet{johnson2015}.

Nitrogen concentrations in continental crust rocks have increased with time. Upper crust concentrations from tills are $66\pm100$\,ppm for the Neoarchean and Palaeoproterozoic and $290\pm165$\,ppm for the Neoproterozoic to present day \citep{Johnson2017}. This change implies net draw-down from the atmosphere. 

Mantle nitrogen is of subducted origin \citep{marty-95}, with a return flux via mantle outgassing. This is evident from noble-gas systematics, specifically the relationship between N and Ar (which has a similar solubility to N). Nitrogen abundance does not correlate with the primordial Ar isotope, \ce{^{36}Ar}, indicating that most primordial nitrogen has been outgassed. Rather, it correlates well with \ce{^{40}Ar}, the radiogenic daughter product of \ce{^{40}K}. This correlation suggests an mantle inventory a few times the size of the atmospheric reservoir \citep{marty-95,md-03}. Given that K is lithophile (incompatible with mantle rock, preferentially entering melt and thus residing in the crust), the association of N with K requires that the mantle N is of subduction origin \citep{marty-95,md-03}. Given the source species is \ce{NH4^+}, biological fixation of atmospheric \ce{N2} is necessarily invoked \citep{Goldblatt2009n2}.

Whilst the size and source of the mantle reservoir is evident, there are presently neither constraints on its age, nor on the balance of N subduction and outgassing through time. Present day rates suggest slow net transfer of N to the mantle \citep{li-bebout-05,lbi-07,Goldblatt2009n2}, with a mantle N lifetime of billions of years \citep{Goldblatt2009n2}. Nitrogen transfer into the subduction zone depends on how much \ce{NH4^+} can substitute into ocean crust, and on how much nitrogen is retained in sediments; both of these may have been higher in the past when the deep ocean was anoxic. Some proportion of the subducted nitrogen returns to the atmosphere through arc volcanoes, depending on the temperature gradient of the subduction zone \citep{bebout-ea-99}. Assessing the nitrogen transfer across subduction zones is challenging because of the need to fully account for geological nitrogen entering the system, and atmospheric contamination makes estimating the gas flux outward hard. Nonetheless, modern studies do call for net subduction \citep{li-bebout-05,lbi-07,Goldblatt2009n2}.

In developing a theory of atmospheric \ce{N2} evolution, primary uncertainties remain. At the end of accretion and differentiation of Earth, was the N in the mantle or atmosphere? How has the rate of N subduction changed through time, and how has it been balanced by mantle outgassing? A first-order understanding of the \ce{N2} evolution, and thus the evolution of atmospheric mass, is only presently being developed: histories of increasing or decreasing atmospheric di-nitrogen, or of local maxima and minima, are all permissible.

\subsubsection{Reactive N species}

Ammonia and nitrous oxide are both potent greenhouse gases, and may have played a role in keeping early Earth warm. Indeed, ammonia is one of the best greenhouse gases, absorbing strongly in the 10\,$\mu{m}$ water vapour window \citep{sm-72,Byrne2014}, and nitrous oxide is of similar strength to \ce{CO2} \citep{Roberson2011,Byrne2014}. They would require concentrations of 0.8\,ppmv and 5\,ppmv, respectively, in a 1\,bar atmosphere to give a 10\wmm{} radiative forcing and counter 20\% of the 50\wmm{} radiative deficit from the faint young Sun \citep{Byrne2014}.

Ammonia was the original proposal to resolve the Faint Young Sun Paradox \citep{sm-72}. However, this was predicated on a highly reducing Archean atmosphere, which is no longer supported. Ammonia is very soluble in seawater, and photochemically unstable \citep{ka-79} so long-term, high concentrations would have been difficult to support, though the presence of an organic haze could enhance the photochemical lifetime of ammonia \citep{sc-97}.  A trace amount could be sustained by a balance of nitrogen fixation followed by ocean outgassing and photochemical destruction. 

Nitrous oxide is the product of partial denitrification. This could be favoured in anoxic and sulphidic oceans, which may have been the norm in the Proterozoic, as copper limitation may have prevented the formation of the nitric oxide synthase (NOS) enzyme which transitions \ce{N2O} to \ce{N2}, possibly giving \ce{N2O} fluxes twenty times the modern rate\citep{buick-07}. However, reduced atmospheric oxygen leads to faster photolysis of \ce{N2O}, making concentrations above 1\,ppmv difficult to support \citep{Roberson2011}.

\section{Oxygen and ozone}

Di-oxygen (\ce{O2}) is the second most abundant gas in the modern atmosphere (21\%), and ozone (\ce{O3}) is a photochemical product of this. The atmospheric evolution of these are interwoven; producing ozone requires a sufficient oxygen inventory, whereas the formation of the stratospheric ozone layer provided photochemical shielding to the troposphere, allowing oxygen concentrations to rise towards modern levels. The evolution of these two gases must, therefore, be discussed together. 

The di-oxygen in Earth's atmosphere is a biological product, formed during oxygen producing (oxygenic) photosynthesis, and it is able to accumulate in the atmosphere only in the absence of consumption of organic matter by aerobic respiration. The oxygen and organic carbon cycles are thus intrinsically linked. Oxygen is highly reactive (it will oxidise volcanic gases, crustal rocks, organic carbon, and so on), so the oxygen evolution must further be considered in context of the global redox budget. 

After introducing the record of redox evolution of Earth, the processes determining oxygen levels are described, so as to place oxygen evolution in the context of the development of these Earth system processes.


\subsection{The history of O}

Oxygen evolution has occurred in a number of steps. There is ample evidence, described below, for a ``Great Oxidation'' of the Earth, a geologically rapid transition from a reducing to an oxidising atmosphere, early in the Palaeoproterozoic. Oxygen levels are inferred to be parts per million or less prior to this and up to a percent afterwards. Proterozoic oxygen levels were likely a fraction of a percent. The next major transition was in the Neoproterozoic, when oxygen reached several percent, and near modern levels were achieved in the Devonian. The Great Oxidation represents by far the most profound change in geochemistry in Earth history, and is the focus here. 

Oxygen oxidises; thus there is an excellent record from which this history has been elucidated. Traditional indicators of oxygen levels use the redox state of minerals in the sedimentary record \citep{h-84}.
Iron is one of the most important of these redox sensitive minerals, with two oxidation states: a reduced species \ce{Fe^{2+}} (ferrous), which is water soluble, and an oxidised species \ce{Fe^{2+}} (ferric), which is insoluble. Two classical oxygen indicators derive from this, and serve as tutorial examples. Redbeds are detrital sedimentary rocks (commonly sandstones), which contain a red ferric oxide cements formed through subaerial alteration in an oxidising atmosphere; they are found since 2.3\,Ga. Banded iron formations are ocean sediments containing alternating layers of chert and either magnetite or haematite. These are formed distant from any iron source, so require long range transport of iron in the ocean. This, in turn, requires a reduced ocean such that iron is soluble. Banded iron formations occur mostly before 2.4\,Ga \citep{h-84, ia-99}. Other classical redox indicators include detrital grains of reduced minerals common in Archean sediments (requiring a reduced atmosphere) and changes in the mineralogy of palaeosols \citep[see ][for a wide ranging discussion]{h-84}. The ensemble of these is sufficient that it has long been possible to identify a change from a reducing to an oxidising atmosphere around the Archean-Proterozoic boundary. 

Twenty-first century work has constrained oxygen evolution via numerous isotope systems. The defining modern indicator of the Great Oxidation comes from sulphur isotopes \citep{Farquar2000}. The sulphur isotope record prior to 2.4\,Ga is characterized by anomalies from the expected mass dependent trend in triple-isotope measurements, so called ``mass independent fractionation'' of sulphur isotopes (MIF-S), whereas this feature is absent in younger sediments. The anomalies are photochemically generated. They require (1) a sufficient flux volcanic sulphur to the atmosphere, (2) the absence of an ozone layer, which would provide photochemical shielding, and (3) a reducing atmosphere, which allows multiple exit pathway for sulphur such that isotopic heterogeneity can be preserved \citep{zcc-06}. The existence of the MIF-S record thus constrains Archean oxygen to a maximum of part per million levels \citep{pk-02,zcc-06}.

The currently prevailing interpretation of Proterozoic oxygen history is (1) the Great Oxidation event in the earliest Palaeoproterozoic, which is coeval with low-latitude glaciation (2) peak levels for a couple of hundred million years in the Palaeoproterozoic (3) a decrease to a fraction of a percent, which is maintained for most of the Proterozoic \citep{Lyons2014}, then (4) a second oxygenation to perhaps 1\% of modern levels in the Neoproterozoic. The exact time is not constrained, but it may well occur coeval with another set of low-latitude glaciations in the Cryogenian. The first large metazoa in the fossil record, which likely required percent \ce{O2}, appear in the Ediacaran \citep{Och2012}The Palaeoproterozoic peak in oxygen is linked to abundant organic carbon burial causing an oxygen source, which is required to explain the Lomagundi carbon isotope excursion. Mesoproterozoic oxygen levels are constrained by a variety of trace metal systems to a fraction of a percent, and imply an anoxic, ferruginous and locally euxinic ocean, for much of the Proterozoic \citep{Lyons2014}

A most remarkable aspect of the 2.4\,Ga Great Oxidation is how late it is. There are numerous lines of evidence for the origin of oxygen producing photosynthesis several hundred million years earlier \citep[see][for a recent discussion]{Lyons2014}. 
The motivating question becomes: why was the rise of atmospheric oxygen delayed following the advent of oxygen production?


\subsection{Sources and sinks of atmospheric oxygen} \label{s-ox}

The oxygen cycle begins with a fast biological cycle, closed either within the biosphere or by atmospheric chemistry.  A slow geologic cycle involves burial and subsequent oxidative weathering of organic carbon. 

It is a common misconception to simply equate atmospheric oxygen to oxygen production by photosynthesis. The problem is, of course, that in the presence of abundant oxygen, the vast majority of the carbon fixed (reduced from \ce{CO2} to organic carbon, and represented schematically as \ce{CH2O}) is oxidised on short timescales by aerobic respiration:
\begin{equation}
\ce{CO2 + H2O <=>[\text{oxygenic photosynthesis}][\text{aerobic respiration}] CH2O + O2} \label{e-oxps}
\end{equation}
That is, the modern oxygen production cycle is largely closed by biology.
A long-term source of oxygen to the atmosphere requires burial of organic carbon in sediments and sedimentary rock, leaving the \ce{O2} in the atmosphere. Oxidation of organic carbon during weathering is, equivalently, a sink for atmospheric oxygen. The reservoir of sedimentary organic carbon is $1\times 10^{21}$\,mol, implying a time integrated oxygen source of $\sim\!30$ times the atmospheric inventory \citep{cc-05}. 

Where high oxygen, aerobic respiration, and closure of the oxygen production by biology is one end member, the other end member is anoxia, wherein fermentation and methanogenesis decompose organic matter and ultimately closure of the cycle is via atmospheric chemistry. 
Net methanogenesis can be represented schematically
\begin{equation}
\ce{CH2O -> 1/2CO2 + 1/2CH4} \label{e-methanogen}
\end{equation}
so the net effect of photosynthetic production and decay of organic matter is 
\begin{equation}
\ce{1/2CO2 + H2O -> O2 + 1/2CH4} \label{e-netpsmethanogen}
\end{equation}
That is: given anoxia, there should be a large and stoichiometrically balanced flux of oxygen and methane from the biosphere to the atmosphere. 
The closure of the system is by net methane oxidation, which is photochemically mediated:
\begin{equation}
\ce{ 1/2CH4 + O2 -> H2O + 1/2CO2} \label{e-atmosox}
\end{equation}
The rate of net methane oxidation has a non-linear dependance on oxygen and ozone levels (described below). High atmospheric methane is expected.

An additional, and important, net source of oxygen is hydrogen escape to space \citep{hd-76}. High in the atmosphere, any hydrogen bearing compounds may be photolized, freeing hydrogen atoms light enough to escape Earth's gravity. For Earth-like, hydrogen-poor, atmospheres the rate limiting step is upward diffusion of hydrogen bearing compounds from the homopause (the level at which the atmosphere ceases to be well mixed, $\sim\!100$\,km for Earth). The escape flux thus depends directly on both the total hydrogen mixing ratio at the homopause and on gravity (the latter because diffusion upwards is a buoyancy flux). Because water condenses at atmospheric temperatures, it is scarce at the homopause, and methane likely contributed the most hydrogen for escape \citep{czm-01}. Schematically hydrogen escape can be represented, 
\begin{eqnarray}
\ce{CH4} &\ce{->}& \ce{ 4H (\textrm{to space}) + C} \\ 
\ce{C + O2} &\ce{->}& \ce{ CO2} \label{e-hesc}
\end{eqnarray}
The oxygen source arises as two moles of oxygen are produced for each mole of methane (Eqns. \ref{e-oxps} to \ref{e-netpsmethanogen}) but only one oxygen is used associated with hydrogen escape (Eqn \ref{e-hesc}). 
Taking a representative methane mixing ratio of 100\,ppmv, for the first half of Earth history, the oxygen flux would be $0.7\times 10^{12}$\,mol\,yr$^{-1}$ and the total oxygen source is $1.6\times 10^{21}$\,mol, equivilant to 50 times the modern atmospheric oxygen inventory \citep{czm-01,cc-05}. 

Major oxygen sinks arise from oxidation of crustal rocks and volcanic gases. \citet{cc-05} estimate the excess of oxidised iron (\ce{Fe^{3+}}) and sulphate in rocks and sediments to be equivalent to a time-integrated oxygen sinks of around $2.5\times10^{21}$\,mol and $0.5\times10^{21}$\,mol respectively, together 100 times the present atmospheric \ce{O2} inventory. 

Mass balance requires that time-intergrated oxygen sources and sinks balance: oxidized reservoirs, atmospheric \ce{O2} and (dominantly) the oxidised crust, are balanced by a mix of burial or organic carbon and hydrogen escape.   

\subsection{Photochemistry, ozone, and a model for the Great Oxidation}

Ozone, \ce{O3}, is a photochemical product of \ce{O2}, so the generation of a stratospheric ozone layer is intimately connected to the rise of oxygen as a major atmospheric gas. Establishment of an ozone layer requires a threshold level of oxygen. Once established, the ozone layer provides photochemical shielding of the troposphere, thus reducing efficiency of methane oxidation and facilitating the accumulation of atmoshperic oxygen. Thus, the formation of the ozone layer is the mechanism of the Great Oxidation \citep{Goldblatt2006}.  

The reaction sequence for ozone production \citep[][CHAPTER-REF Ozone and Stratospheric Chemistry]{Chapman1930} is
\begin{eqnarray}
\ce{O2} + h\nu(\lambda < 240)\,\text{nm} & \ce{->} & \ce{O + O}  \label{e-o2split}\\
\ce{O + O2 + M} & \ce{->} & \ce{O3 + M}\\
\ce{O3} + h\nu(\lambda < 320\,\text{nm})& \ce{->} & \ce{O2 + O}\\ 
\ce{O3 + O} & \ce{->} & \ce{2O2} 
\end{eqnarray}
Ozone production requires sufficient oxygen levels to produce odd oxygen via Eq \ref{e-o2split}. In photochemical models, some ozone production can be seen at $\sim\!1$\,ppm \ce{O2}, whereas a modern-like ozone layer forms at 1\% of present oxygen levels \citep{kd-80}.

Methane oxidation rates, as with much atmospheric chemistry, depends on the availability of \ce{OH^-} radicals. The main source of \ce{OH^-} is photolysis of water vapour which, given the decrease in atmospheric temperature with altitude, is found dominantly in the troposphere. The existence of the ozone layer and consequent UV shielding of the troposphere reduces tropospheric \ce{OH^-} availability and slows down much of the atmospheric chemistry. 

Thus, with an ozone layer, the effective rate constant for methane oxidation is lower, which drastically changes the dynamics of oxygen. Given the same stoichiometically balanced flux of methane and oxygen from the biosphere, the oxygen level necessarily should increase so that the photochemical sink balances the source. Equivalently, the lifetime of oxygen is extended. 

Two rather distinct stable states of oxygen exist: a low oxygen stable state with around a part per million oxygen, and a high oxygen stable state of near a percent. There is a very non-linear transition between these: the Great Oxidation \citep{Goldblatt2006}. 


\section{Carbon dioxide}

\begin{table}
\begin{center}
\begin{tabular}{ll}
\hline
\textbf{Reservoir}  & \textbf{Size ($10^{12}$\,kg\,C)} \\
\hline
\textbf{Biomass (live)}   & $\mathbf{600}$ \\
   \quad Land & $600$ \\
   \quad Ocean & $6$  \\
 \textbf{Soils and sediment}   & \textbf{4100} \\
   \quad Organic & 1600 \\
   \quad Inorganic (carbonate) & 2500  \\  
\textbf{Ocean (inorganic)} &  \textbf{39,000} \\
\textbf{Atmosphere} &  $\mathbf{870}$ \\  
  \quad Methane & 1.7\\
  \quad Carbon dioxide & 868 \\
\textbf{Methane Hydrates} &  $\mathbf{2400\pm1200}$ \\ 
\textbf{Sedimentary rock} & $\mathbf{7.5 \times 10^7}$\\
   \quad Organic & $1.5 \times 10^7$\\
   \quad Inorganic (carbonate) & $6 \times 10^7$\\
\textbf{Mantle} & $\mathbf{(7.0\pm 1.7)\times 10^8}$\\
   \quad MORB Source & $(3.8\pm 1.5) \times 10^8 $\\
   \quad OIB Source & $(3.2\pm 0.8) \times 10^8 $\\
\textbf{TOTAL} & $\mathbf{\sim\!7.8\times 10^8}$\\   
  \hline 
\end{tabular}
\end{center}
\caption{Global carbon budget. Atmospheric \ce{CO2} was 403\,ppmv at the time of writing. Mantle reservoirs are from \citet{Anderson2017,LeVoyer2017}, but method-dependent uncertainties are a factor of a few. Other reservoirs from \citet{Kump2009, Wallmann2012} (error for rock sedimentary reservoirs is not stated, but will be non-trivial).}\label{t-carbon}
\end{table}

Carbon has a variety of valence states; in the atmosphere carbon dioxide (\ce{CO2}) is the main oxidised species and methane (\ce{CH4}) the main reduced species. Carbon dioxide has likely been the dominant greenhouse gas throughout Earth history, with methane a popular candidate for second place. In terms of both budget and processes, there is a fundamental split between the cycles of inorganinc (oxidised) and organic (reduced) carbon. The main link between these is oxygenic photosynthesis, reversed in oxidising conditions by aerobic respiration (Equation \ref{e-oxps})
or in anaerobic conditions by methanogenesis and subsequent atmospheric methane oxidation (Equations \ref{e-methanogen} and \ref{e-atmosox}). The link between organic carbon and oxygen is so strong that atmospheric methane and the organic carbon budget are considered alongside oxygen (Section \ref{s-ox}). 

The  carbon budget is given in Table \ref{t-carbon}. Carbon dioxide is a minor species today, but the vast size of geological reservoirs of inorganic carbon make a much higher inventory in the deep past easy to motivate. Indeed, the carbon dioxide inventory of Venus' atmosphere is essentially equivalent to the geological C on Earth (see section \ref{s-comparative}). On Earth, the inorganic carbon (limestone) reservoirs are largely biogenic. On shorter timescales, the intermediate-sized ocean reservoirs provide rapidly-exchangeable carbon; this has dominated \ce{CO2} change through Quaternary glacial cycles. 

Atmospheric carbon dioxide is controlled by a hierarchy of processes, depending on the timescale. On sub-annual timescales, higher rates of photosynthesis during the summer and respiration during the winter, especially from northern hemisphere land biota, give a hemispheric annual cycle of 10\,ppm. On annual to centennial timescales, the dominant processes changing \ce{CO2} now are anthropogenic fossil fuel burning and cement production (CHAPTER-REF ?), which are increasing \ce{CO2} by 2 ppm per year\footnote{\url{https://scripps.ucsd.edu/programs/keelingcurve/}}. More generally, annual to millennial control on \ce{CO2} is via ocean carbonate chemistry (changes in the speciation of inorganic carbon in the ocean) \citep{Ridgwell2005, ZeebeWolfGladrow}. This, in turn, is very heavily modulated by life and, on $\gtrsim 10^5$\,yr timescales, by geological activity. Presence of widespread and cumulatively vast deposits of biogenic calcium carbonate rock throughout Earth history provide a \emph{prima facie} case for the combined role of life, ocean chemistry and geology in controlling \ce{CO2}. 

\subsection{Control of \ce{CO2} by ocean chemistry}

Carbon dioxide readily dissolves in sea water. The equilibrium dissolved carbon dioxide is given by Henry's Law
\begin{equation}
 [\ce{CO2}] = k_H \ce{CO2}
\end{equation}
There is a similar amount of dissolved molecular \ce{CO2} in the ocean as atmospheric \ce{CO2}.  Once in solution, however, the total amount of dissolved inorganic carbon (DIC) partitions between \ce{CO2}, \ce{H2CO3} (carbonic acid), \ce{HCO3-} (bicarbonate ion) and \ce{CO3^{2-}} (carbonate ion). For Henry's law, only the dissolved \ce{CO2} matters (i.e. that is all the atmosphere ``sees'') so the sum of these species, the DIC, is 50 times larger than dissolved \ce{CO2} alone.

The equations for the partitioning are:
\begin{eqnarray}
\ce{CO2} + \ce{H2O} &\ce{<=>} & \ce{H2CO3}  \\  \label{e-carbpartition1}
\ce{H2CO3} & \ce{<=>} & \ce{H^+} + \ce{HCO3^-} \\ \label{e-carbpartition}
\ce{HCO3^-} & \ce{<=>} & \ce{H^+} + \ce{CO3^{2-}} \\ \label{e-carbpartition2}
\end{eqnarray}
Which give 3 equations with 5 unknowns. Two more constraints are needed. First is that DIC is a conserved quantity (conservation of mass of carbon atoms):
\begin{equation}
\text{DIC} = [\ce{CO2}] + [\ce{H2CO3}] + [\ce{HCO3^-}] + [\ce{CO3^{2-}}]
\end{equation}
The second conserved quantity is alkalinity, which accounts for conservation of charge. Ions in the ocean may, conceptually, be split into two groups. First, those from dissolution of salts and strong acids; negative changer from \ce{Cl-} and others \textit{almost} balances positive charge from \ce{Na+}, \ce{Mg^{2+}}, \ce{Ca^{2+}} and others. These are referred to as conservative ions. Second, alkalinity is the net negative charge provided by species which exchange protons at $pH>4.5$, which balances the net charge from conservative ions. For instructive purposes (in terms of the equations above), we can define alkalinity
\begin{equation}
\text{Alk} = [\ce{HCO3-}] + [\ce{CO3^2-}] \quad(\ldots - [\ce{H+}] + \ldots ) \label{e-alk}
\end{equation}
The first two terms on the right hand side are the ``carbonate alkalinity'', which dominate. The bracketed term is small, but included to show, conceptually, how the Equations \ref{e-carbpartition1}--\ref{e-carbpartition2} are balanced. In the real ocean, borate contributes at percent levels and there are additional minor contributing species. \citet{ZeebeWolfGladrow} provides and excellent textbook description of these processes. 

\begin{figure*}
\begin{center}
\includegraphics[width=5cm]{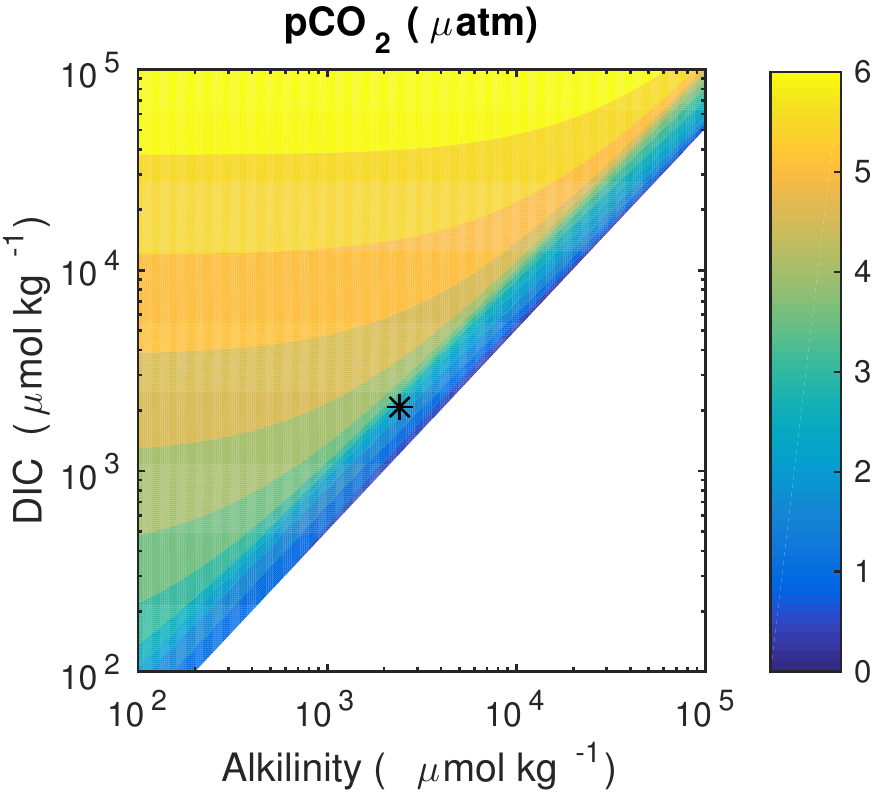} \includegraphics[width=5cm]{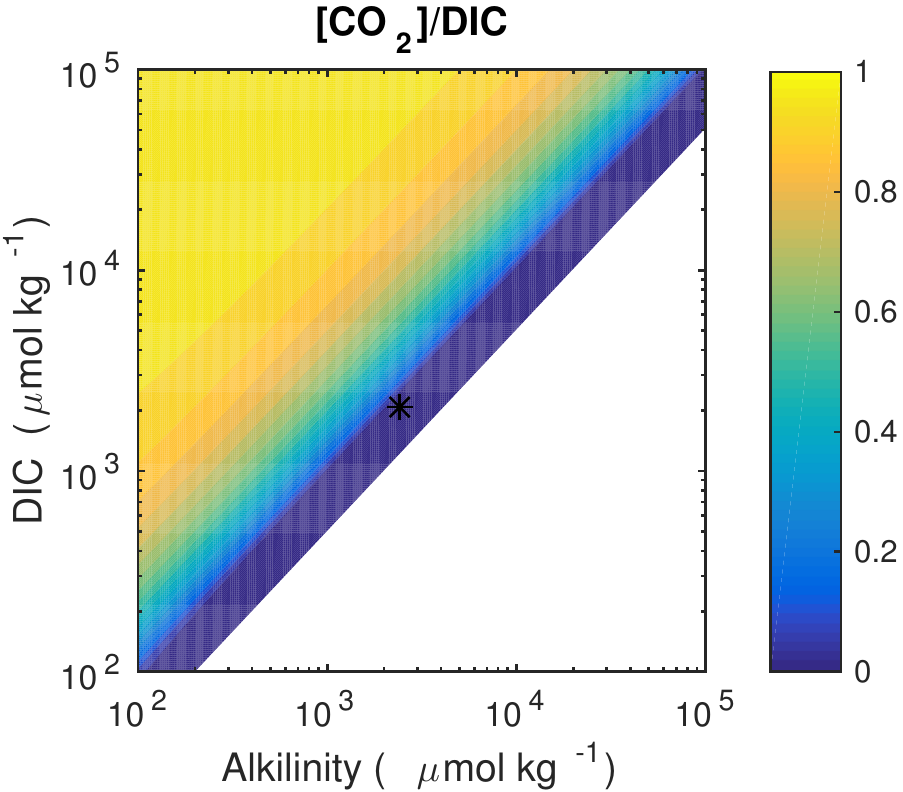}

\includegraphics[width=5cm]{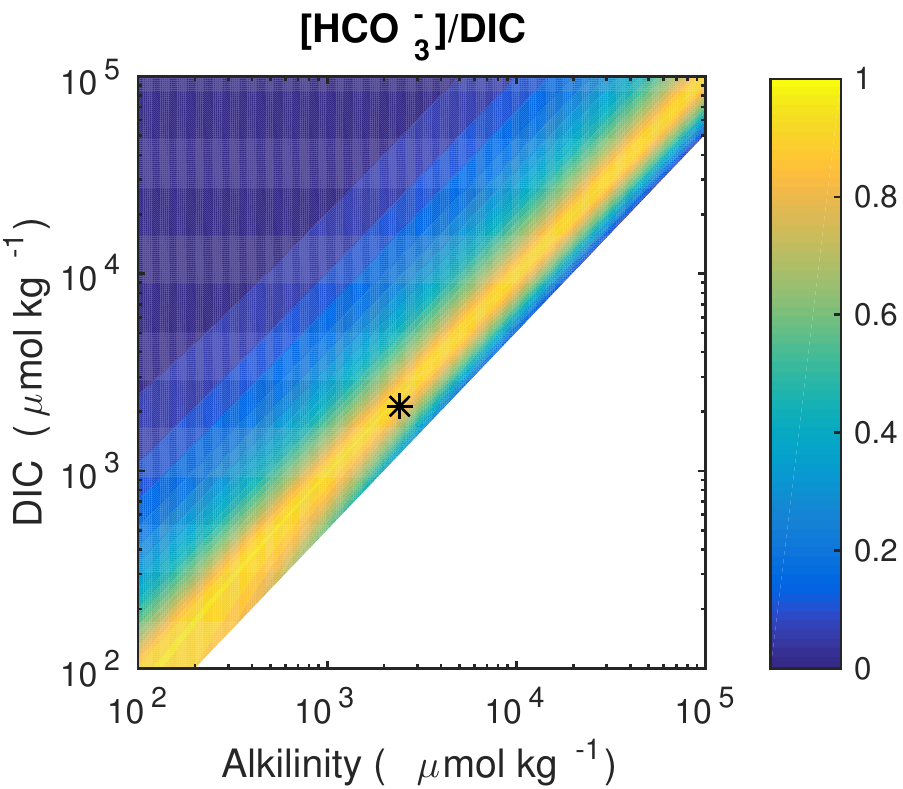} \includegraphics[width=5cm]{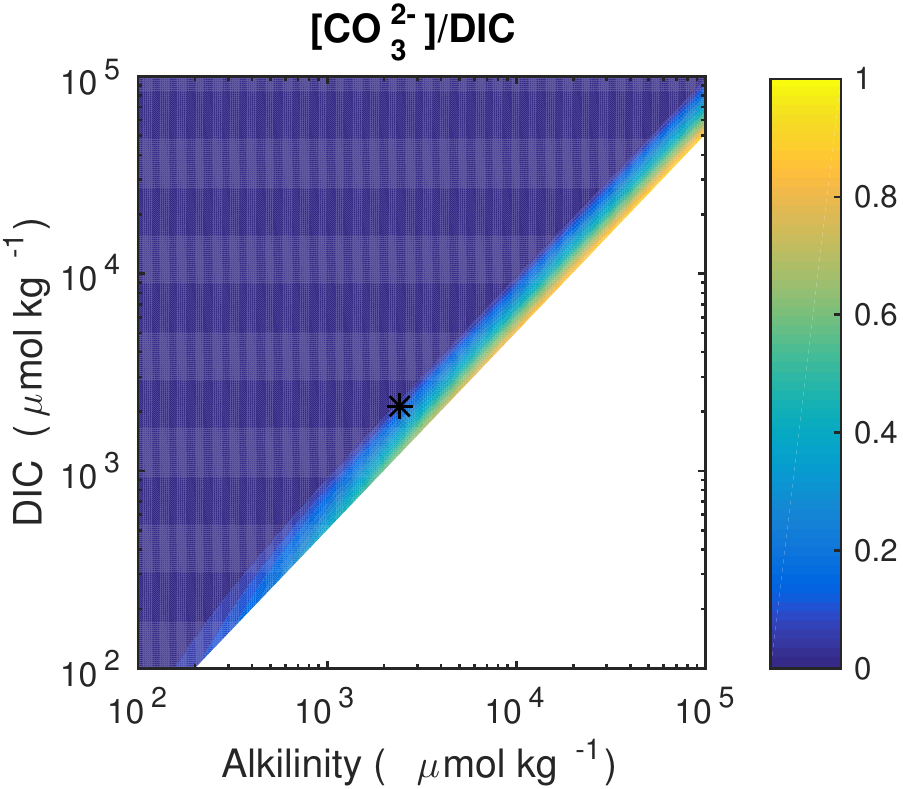}
\end{center}
\caption{Atmospheric carbon dioxide and dissolved species concentrations as a function of DIC and Alk. There is a tendency to think of the total amount of carbon in the atmosphere-ocean determining the strength of the \ce{CO2} greenhouse, but quite clearly alkalinity exerts equal leverage. Figures made with the Zeebe code \citep{ZeebeWolfGladrow}} \label{f-co2sys}
\end{figure*}

Specifying any two parameters in the carbonate system (Equations \ref{e-carbpartition1} -- \ref{e-alk}) is sufficient to determine the whole system. DIC and alkalinity are the conserved quantities and have clear physical meaning (total carbon and charge balance), so make most sense to use. 
Figure \ref{f-co2sys} shows the partitioning of carbon between different reservoirs. What should be immediately obvious is that $p$\ce{CO2}---hence the contribution of inorganic carbon to the greenhouse effect---cannot be determined directly from the DIC: knowledge of alkalinity in needed. That is, ocean chemistry is fundamental to the abundance of the dominant non-condensible greenhouse gas in Earth's atmosphere.

\subsection{Calcium carbonate formation}

Ocean chemistry is linked to sediments and sedimentary rock via the formation and dissolution of \ce{CaCO3}:
\begin{equation}
\ce{Ca^{2+} + CO3^{2-} <=> CaCO3} \label{e-calcite}
\end{equation}
This has important consequences for p\ce{CO2} and climate on millennial and longer timescales. 

The position of solution with respect to equilibrium in Equation \ref{e-calcite} can be expressed as the calcite saturation, $\Omega_\text{calc}$
\begin{equation}
\Omega = \frac{\ce{[Ca^{2+}][CO3^{2-}]}}{k_\text{sp}}
\end{equation}
As an approximation, carbonates will decompose when $\Omega_\text{calc}<1$ whereas formation requires supersaturation, perhaps  $\Omega_\text{calc}\approx30$ for chemical precipitation (also referred to as authigenic formation) or $\Omega_\text{calc}$ of 1 to 3 for biologically mediated formation. Variation in [\ce{CO3^{2-}}] dominates over variation in [\ce{Ca^{2+}}], hence the ratio of alkalinity to DIC has a controlling influence on \ce{CaCO3} precipitation: a high alkalinity to DIC ratio gives high [\ce{CO3^{2-}}] and therefore high $\Omega_\text{calc}$.

Calcium carbonate formation removes two units of alkalinity for each unit of DIC. The immediate effect, therefore, is to shift the partitioning of DIC away from [\ce{CO3^{2-}}] and towards [\ce{CO2}], consequently reducing the system's propensity to precipitate more calcite and increasing atmospheric p\ce{CO2}. On longer timescales, however, it is more important to think of this process as removing inorganic carbon from the atmosphere-ocean system, to be discussed further below.

Carbonate sediments are ubiquitous throughout the geological record. Their relatively continual formation places an empirical constraint on $\Omega_\text{calc}$, implying that there have only been minor variations in the ratio of alkalinity to DIC through most of Earth history \citep{Walker1983}. The fundamental question of what operates this control remains somewhat open. The amount of limestone (\ce{CaCO3} in sedimentary rock) is vast:  $5\times10^{21}$\,mol (Table \ref{t-carbon}). Were this in the atmosphere, Earth would have a 42\,bar \ce{CO2} atmosphere and a surface far too hot for life. The role of limestone formation---which is biologically controlled on Earth---in maintaining habitable conditions is evident. 

\subsection{Geological control of alkalinity: the carbonate-silicate thermostat}

An ocean of pure water would not mediate carbonate deposition. Rather fortunately in this regard, Earth's ocean is impure, containing dissolved ions from the chemical weathering of crustal rocks. As alkalinity balances the over-abundance of cations amongst these salts, the cation flux determines long term rates of carbonate deposition. 

Weathering of crustal silicate rocks can be represented with the following series of reactions, using simplified stoichieometry. Atmospheric \ce{CO2} dissolves and hydrates to become carbonic acid, which reacts with silicate rock (represented \ce{CaSiO3}):
\begin{eqnarray}
2(\ce{CO2 + H2O} & \ce{->} & \ce{H2CO3})  \\
\ce{CaSiO3 + 2H2CO3} &  \ce{->} & \ce{Ca^{2+} + 2HCO3^{-} + SiO2 + H2O}  
\end{eqnarray}
Two units of \ce{CO2} are removed from the atmosphere and two units of both DIC and Alk are added to the ocean for each unit of silicate weathered. Given that \ce{CO2} exchange between atmosphere an ocean is rapid, DIC will be removed by gas flux to the atmosphere. Increasing Alk  then dominates in the ocean, and a shift in DIC from bicarbonate to carbonate ions causes carbonate deposition,
\begin{eqnarray}
\ce{2HCO3^{-}} & \ce{->} & \ce{CO3^{2-} + CO2 + H2O} \\
\ce{Ca^{2+} + CO3^{2-}} & \ce{->} & \ce{CaCO3}
\end{eqnarray}
removing two units of both alkalinity and DIC from the ocean. The net stoicihiometry of all of these is
\begin{eqnarray}
\ce{2CO2 + 2H2O + CaSiO3} & \ce{->} & \ce{SiO2 + CO2 + 2H2O + CaCO3} \\ \nonumber
\ce{CO2 + CaSiO3} & \ce{->} & \ce{SiO2  + CaCO3}
\end{eqnarray}
This system is known as the Urey reactions \citep{Urey1952}. Thus, silicate weathering followed by carbonate deposition leads to removal of inorganic carbon from the atmosphere-ocean system. 

In a seminal paper, \citet{whak-81} described a negative feedback on the long-term climate of Earth. Carbon dioxide is a greenhouse gas, so higher atmosphere-ocean carbon should result in a warmer planet and faster reaction rates. Likewise, a warmer planet would have higher rainfall which would enhance weathering. Warming would lead to carbon dioxide draw-down, providing a negative feedback on global temperatures. The WHAK feedback (known for Walker, Hays and Kasting), remains the dominant paradigm for temperature regulation on Earth. 

Its utility notwithstanding, there are various nuances and complications of the behaviour of this system, such that one should look beyond the idea of a set point temperature toward which Earth is regulated. These include system dynamics, the mechanics of weathering, and the role of life. 

Any dynamical model (there are many) will ultimately balance a volcanic \ce{CO2} source with weathering-motivated carbonate deposition. Decreasing outgassing rate or increasing the reaction coefficient for weathering will then give a colder steady state; decrease these far enough and glaciation will ensue. Cold temperatures curtail weathering, atmospheric \ce{CO2} accumulates and Earth warms; thus limit-cycle behaviour can emerge \citep{mills2011,Abbot2016}. 

Weathering does not occur uniformly. Rates are highest where there is fresh volcanic rock, young mountains, and warm and wet conditions. 50\% percent of global chemical weathering occurs on 9\% of the surface \citep{Hartmann2009}. 
Broadly, felsic rock (e.g. granite) is is less susceptible to weathering than mafic rock (e.g. basalt) \citep{Ibarra2016}, so weathering can be dominated by fresh exposures of mafic, volcanic rock. Today, basalts are responsible for 30---35\% of the global \ce{CO2} sink by weathering, dominated by tropical volcanic island arcs \citep{Dessert2003}. In the past, migration of large igneous provinces (flood basalts) through the humid tropics may have been responsible for major cooling; hypothesis attribute Cenozoic cooling \citep{Kent2013} and the Neoproterozoic snowball \citep{Godderis2003} to this mechanism, for example. 
For much of the continental interiors, where there is little variation in elevation, chemical weathering is limited by the availability of fresh rock and is not temperature dependence; only in mountainous areas, where exposure of fresh rock is rapid, are weathering rates temperature dependent \citep{west-ea-05}. 

Life is deeply implicated in the control of alkalinity, and therefore atmospheric \ce{CO2}. In soils, p\ce{CO2} is elevated above atmospheric levels by respiration \citep{blag-83}. Furthermore, fungal hyphae and plant roots  penetrate rock, excreting organic acids which enhance weathering. The evolution of lichens (probably in Neoproterozoic time), a symbiosios of algae which provide a photosynthetic source of carbon and energy, and fungus which extract nutrients, may have marked the beginning of very efficient, photosynthetically driven, enhancement of the weathering \citep{Berner1992}. The evolution of non-vascular plants (mosses) in the Ordovician, and vascular plants in the Devonian further enhanced alkalinity supply by a factor of 2 to 10 \citep{Lenton2012}. Enhanced weathering has been linked to various glaciations \citep[e.g.][]{Berner1997,Royer2006,Lenton2012} In the ocean, organisms precipitate \ce{CaCO3} at much lower $\Omega$ than authigenic precipitation \citep{ZeebeWolfGladrow}. The biota thus cause atmospheric \ce{CO2} to be lower than it would be on an otherwise equivalent but sterile planet.

\subsection{Proxy constraints on, and history of, atmospheric \ce{CO2} }

With \ce{CO2} as the main non-condensible greenhouse gas and a fainter Sun in the past, simple theoretical treatment of silicate-weathering feedback would call for p\ce{CO2} to be high in the Archean and  decrease over time to modern values. Since the proposal of high \ce{CO2} as the resolution to the faint young Sun problem, there has been a consistent mis-match between climate model estimates of how much \ce{CO2} should be required, and proxy estimates of how much is in the atmosphere. 

For the Neoarchean and Proterozoic, palaeosol constraints have been used for some decades. Soils are necessarily in contact with the atmosphere, so can be used as an atmospheric proxy. The absence of siderite, can be used to exclude very high p\ce{CO2} \citep{rkh-95}. Models can be used to quantify p\ce{CO2}, e.g. \citet{Sheldon2006} suggests $23^{\times 3}_{\div 3}$ times the present level, but key parameters (e.g. soil formation rate) are very poorly constrained, making such model interpretations very challenging. 

Recently, calcium isotopes have been used to more directly constrain the carbonate system via [Ca]/Alk; still, this does not uniquely constrain p\ce{CO2}, and a supplemental assumption about pH is necessary. A nominal assumption of pH of 7 would give around 30,000\,ppmv \ce{CO2}. However, if another carobonate system parameter can be determined independently, then an actual constraint on p\ce{CO2} will be available \citep{Blattler2017}.

For the Phanerozoic, a variety of proxies exist and give a somewhat consistent picture: p\ce{CO2}  decreases from around 5000\,ppmv in the Silurian to less than 1000\,ppmv in the Carboniferous, then varies between a few hundred and a few thousand ppmv until the present day \citep{Royer2006}.
 
\section{Water}



Water is unique amongst Earth's atmospheric gases in that the control on its abundance is physical: equilibrium between vapour and a large condensed reservoir known as ``the ocean''. 
This physical control is via the saturation vapour pressure: the amount of vapour in equilibrium with a condensed reservoir depends exponentially on temperature. At the mean surface temperature of $\overline{T_s} = 15$\dc\ the saturation vapour pressure of water is 1.7kPa, implying that water comprises just under 2\% of the atmosphere. In reality, it is somewhat less: the atmosphere cools with height, so the saturation vapour pressure decreases strongly with altitude, and in general the atmosphere is sub-saturated with water vapour. Spatial variations are important: Earth's warm tropics are very moist and the cold polar regions quite dry; water vapour is not well-mixed. 

Water vapour content has changed by several orders of magnitude, co-varying with temperature. In times when Earth was much colder, for example during Palaeo- and Neoproterozoic Snowball Earth episodes with $\overline{T_s} \approx -40$\dc, the corresponding saturation vapour pressure was 100 times lower than today. During hyperthermal periods, such at the Cretaceous warm period with $\overline{T_s} \approx 30$\dc, the corresponding saturation vapour pressure was 2.5 times higher than today. Over these range of temperatures, which characterize most of Earth history, the expectation is that water vapour has varied by more than two orders of magnitude, between a trace to minor constituent of the atmosphere. 

Based on these arguments, three qualitatively different atmospheric water inventories, corresponding to different climate states, can be inferred to have existed through Earth history: cold and dry, with trace water in equilibrium with an ice surface (snowball Earth); intermediate climate with strong meridional gradients of temperature and water vapour, so that water is a minor (percent level) constituent in the tropics but trace at the poles (e.g. present day); and warm and moist in equilibrium with a warm ocean, small meridional temperature gradients and water vapour as a minor constituent throughout. 

When water becomes a major constituent, there are further qualitative changes to system behaviour:  transition to a hot and water-dominated atmosphere occurs above a threshold solar constant. That is, a liquid water ocean is stable only up to a limit on the energy supplied to the planet, after which a ``runaway greenhouse'' occurs, the ocean evaporates and the atmosphere becomes water dominated \citep{simpson-27,komabayashi-67,Ingersoll1969}. Given a water vapour column of 30\,kg\,m$^{-2}$, corresponding to a saturated atmosphere with surface temperature of 300\,K, the atmosphere is optically thick at all thermal wavelengths. Consequently, only radiation from the atmosphere can reach space, and none from the surface, limiting outgoing thermal radiation. If more energy is absorbed by the planet, there will be a non-linear transition of the atmosphere as the entire ocean evaporates \citep[see reviews by][]{nakajima-ea-92, Goldblatt2012}. Modern calculations of the radiation limit in clear-sky conditions give 283\wmm, which corresponds to the sunlight that would be absorbed by a cloud-free planet at Earth's orbit \citep{Goldblatt2013}. Clouds are net reflectors, so increase the threshold insolation to at least 1.1 times the present solar constant \citep{Leconte2013,Wolf2014,Popp2016}. As the solar constant increases, Earth will transition to a water-dominated atmosphere in around a billion years. Likewise, Venus receives 1.9 times the sunlight that Earth does, and the composition of the Venusian atmosphere is consistent with a post-runaway state. Whether Venus had oceans which evaporated at some point in the past, or never condensed any, is model dependent \citep{Hamano2013}. 

Additionally, impact energy could be sufficient to evaporate the oceans and give a water dominated atmosphere. This undoubtedly occurred during accretion. During the end-Hadean late heavy bombardment, statistics indicate that there would have been zero to four impacts sufficient to evaporate the oceans \citep{zahnle-ea-07}. 
The timescale of an impact-generated steam atmosphere would be a few thousand years \citep{zahnle-ea-07}. 
Likewise, early in the evolution of a planetary system, tidal heating of the planet could induce a runaway greenhouse \citep{Barnes2013}.


\section{Discussion}
\subsection{Atmospheric homoeostasis, Vernadsky's ``biosphere'', and the Gaia hypothesis}

The most remarkable observation of Earth history is the continual lineage of a single genesis of life spanning four billion years. Indeed, deep in the Archean record, the evidence for life is in general commensurate with the maximum that could be expected, given the preservation of the sedimentary record \citep{ns-01}. Life requires temperature to be in a somewhat limited range, by necessity maintained for the entire length of the record. Three alternate explanations for this long-term homeostasis may be offered: luck, abiological regulation, or an explanation based on the action of life itself. 

Luck is a somewhat irrefutable option. Further disentanglement becomes difficult due to observer bias: our position as observers of this history is contingent on history itself: conditions must have been such that the long evolution to organisms with the science of geology and printing presses to make encyclopaedia could occur. Our ability to pose the question requires continual habitability, so the observation itself is bias; such is the \emph{weak anthropic principle} \citep{wat-99}.

Abiological regulation is supported by evidence of chemical feedback processes contributing to climate regulation, with the negative feedback on temperature in the silicate weathering and carbonate deposition cycle \citep{whak-81} the seminal contribution. A purely abiological model would have these geochemical mechanisms regulate temperature to a level at which life is plausible, allowing life to adapt to the environment. There is little doubt that  this represents part of the explanation, but the entanglement of life in these geochemical mechanisms, enhancing both weathering and carbonate deposition, makes isolation to only abiological processes rather difficult. Life and non-life processes are deeply entwined on Earth.

Biological control was a part of Vernadsky's original enunciation of the biosphere. \citet{vernadsky-26} described the biosphere as composing both living and non-living parts, the atmosphere being the type example of the latter. He saw life as the dominant geological force, and that the planetary scale influence of life has increased with time---i.e. that the biota controlled the atmospheric composition, and that the control has become stronger has evolved to increasing complexity and dominance. 
Written in Russian and French, Vernadsky's original and visionary work on the biosphere was largely lost to anglophone science until David Langmuir's 1970's translation circulated in the late twentieth century, and was eventually published in 1998 \citep[but see also][]{vernadsky-45}.

In western science, biological regulation was proposed by \citet{lov-72} with the \emph{Gaia hypothesis}: ``homoeostasis by and for the biosphere''. A modern statement of this is ``Organisms and their environment evolve as a single, self-regulating system'' \citep{lov-03}. Thus, not only are organisms selected for their environmental fitness, but are selected for their ability to modify the environment in beneficial ways \citep{Lenton1998}.

It is plain that biology is deeply and intimately involved in the control of Earth's atmospheric composition. The question of whether this has directionality---whether the biota regulates---is probably the single most important open question in the study of atmospheric evolution of Earth. Significant theoretical development is needed and, ultimately, experiments to detect life on planets outside the solar system may provide the sample size to resolve the question empirically. 

\subsection{Comparative planetology}\label{s-comparative}

Science progresses best with a sample size greater than one. Thus, as in seeking to understand the evolution of our own atmosphere, much is to be learnt from looking outward. In our solar system, primary comparisons are to our neighbours, Venus and Mars and further information comes from including the giant planet moons (nomenclature and orbit type---planet, moon, dwarf planet---do not affect the geophysics). 

The first, and obvious issue, is whether or not a planet actually has an atmosphere. Earth, Venus, Mars and Titan do, whereas Ganymede, Calisto, Io, Mercury and many smaller bodies do not. Presence of an atmosphere may be empirically determined via the ``cosmic shoreline'': a power law relating threshold insolation to escape velocity to the fourth power. Stellar energy drives atmospheric loss, whereas planetary mass holds the atmosphere down \citep{Zahnle2017}. 

Venus is the most Earth-like planet, with similar size and bulk composition. Venusian atmosphere has a 90\,bar surface pressure (Earth is 1\,bar), with a composition of 97.5\% \ce{CO2} and 3.5\% \ce{N2} and a surface temperature of 700\,K. There is no ocean and water is only a trace atmospheric constituent. 

Venus' current atmosphere is understood on the basis that it experienced a runaway greenhouse in the past \citep{Walker1975}. The present insolation at Venus' orbit is well in excess of the runaway greenhouse limit, so if a Venusian ocean ever condensed after accretion, it later evaporated. Much of the hydrogen from the water was lost to space, evidenced in strong enrichment of D/H relative to Earth \citep{donahue-ea-82}. The timescale for loss of an Earth-size ocean via hydrodynamic escape is a few hundred million years \citep{wdw-81}. Accumulation of oxygen, however, may throttle hydrogen escape \citep{Wordsworth2014}, and extensive hydration of the surface rocks is another potential sink for water \citep{Matsui1986}. 

The carbon reservoir in Venus' atmosphere is of comparable size to the carbon stored in carbonate and organic carbon rocks on Earth. If there was an ocean on Venus, carbonates would have most likely been deposited, but these would have thermally decomposed during the runaway greenhouse (the equivalent process is used in industrial cement manufacture in lime kilns on Earth). 

Venus gives two primary lessons that pertain to Earth. First, is contingency in planetary evolution. Venus' present state is only possible to understand through the presence of atmospheric water in the past, even though it is absent today. Second, is Earth's future: solar constant increases with time, so Earth should reach the runaway greenhouse threshold in a billion years. Venus' past informs us directly or what Earth's future: the end of the world will be a runaway greenhouse, lime-kiln, inferno. 

Mars, by contrast, has a rather feeble atmosphere: 0.006\,bar of \ce{CO2}, supporting a mean surface temperature around 220\,K. Mars' mass and insolation are such that it is only marginally on the atmospheric side of the cosmic shoreline. Yet, there is abundant evidence in Mars' geological record for liquid water in the first billion years of the solar system, despite a dimmer Sun. It is not obvious whether early Mars was always warm, or only episodically so \citep{Wordsworth2016}, but almost any solution requires both a thicker atmosphere and much stronger greenhouse effect in the past.  

Mars and Venus together provide two primary lessons pertinent to Earth. First, that major variations in planetary atmosphere mass and composition, and thus climate, through time are the norm for terrestrial planets. Long term consistency of any of these should be seen as an exceptional result, requiring exceptional evidence and explanation. 
Second, both have \ce{CO2} dominated atmospheres which can be explained by photochemical equilibrium without any biology, in stark contrast to Earth, whose atmospheric composition is to a large extent a biological construct. Life's alteration of Earth is of first order importance.


\section{Outlook}

Humanity is in the midst of a great age of discovery, as planets around other stars are discovered and characterised: ``the age of exoplanets''. Life has had a dominant role in altering the atmospheric composition of Earth, distinguishing us significantly from uninhabited planets. The composition of exoplanet atmospheres will soon be able to be determined remotely via spectroscopy, making \emph{atmospheric analysis} the viable method of detecting life on other planets \emph{via their influence on the atmosphere} \citep{lov-65,hl-67,sagan-ea-93} . 
Life detection will require the detection of atmospheric disequilibrium and the exclusion of any abiotic cause. There is, however, no simple formula for what the composition of an atmosphere should be, so determining whether an observed atmosphere represents a living planet will only be achievable through deep understanding of the processes determining atmospheric composition. Understanding Earth's atmospheric evolution is thus fundamental to learning our place in the universe. 

\bibliographystyle{plainnat}
{\footnotesize
\bibliography{}
}

\end{document}